\documentclass[prl,twocolumn,showpacs,groupedaddress]{revtex4}

\usepackage{graphicx}

\begin{document}

\bibliographystyle{apsrev}

\title{Evidence for Magnetic Field Induced Changes of the Phase
of Tunneling States: Spontaneous Echoes in (KBr)$_{1-x}$(KCN)$_x$
in Magnetic Fields}

\author{C. Enss and S. Ludwig}
\affiliation{Kirchhoff-Institut f\"ur Physik, Universit\"at
Heidelberg, Albert-Ueberle-Str.\ 3--5, 69120 Heidelberg, Germany}

\date{\today}

\begin{abstract}
Recently, it has been discovered that in contrast to expectations
the low-temperature dielectric properties of some multi-component
glasses depend strongly on magnetic fields. In particular, the
low-frequency dielectric susceptibility and the amplitude of
coherent polarization echoes show striking non-monotonic magnetic
field dependencies. The low-temperature dielectric response of these
materials is governed by atomic tunneling systems. We now have
investigated the coherent properties of tunneling states in a
crystalline host in magnetic fields up to 230$\,$mT. Two-pulse echo
experiments have been performed on a KBr crystal containing about
7.5\% CN$^-$. Like in glasses, but perhaps even more surprising in
the case of a crystalline system, we observe a very strong magnetic
field dependence of the echo amplitude. Moreover, for the first time
we have direct evidence that magnetic fields change the phase of
coherent tunneling systems in a well-defined way. We present the
data and discuss the possible origin of this intriguing effect.

\end{abstract}


\maketitle

\narrowtext The low-temperature properties of disordered materials
like amorphous solids and crystals containing certain point defects
are governed by atomic tunneling systems \cite{Esq98,Hun00}. In
glasses such systems arise from the tunneling motion of single atoms
or small groups of atoms between energetically almost equivalent
positions separated by a potential well. In the so-called tunneling
model it is assumed that these atomic tunneling systems can be
approximated by particles moving in a double-well potential
consisting of two adjacent harmonic wells \cite{And72,Phi72}. The
tunneling states are characterized by two parameters, the tunnel
probability $\exp(-\lambda)$ and the asymmetry energy ${\it\Delta}$.
The latter is given by the difference in depth of the two wells. As
a consequence of the irregular structure of glasses, these two
parameters are widely distributed. According to the tunneling model
the asymmetry energy~${\it\Delta}$ and the tunnel
parameter~$\lambda$ are independent of each other and uniformly
distributed as $P(\lambda,{\it\Delta})\,{\rm d}\lambda\,{\rm
d}{\it\Delta} = \overline{P}\,{\rm d}\lambda\,{\rm d}{\it\Delta}$,
providing a description of the low-temperature properties of glasses
on a phenomenological basis.

In contrast to the situation in amorphous solids where the
microscopic nature of the tunneling states is hitherto unknown, in
crystals with certain point defects the tunneling states are
well-defined and can be described on a microscopic basis
\cite{Nar70,Bri75,Wue97}. As prominent examples we mention KCl
containing Li$^+$, and KBr with CN$^-$. In a certain range of
concentrations the system (KBr)$_{1-x}$(KCN)$_x$ is often referred
to as model system for the tunneling states in glasses, because it
exhibits great similarities of its low-temperature properties with
those of amorphous materials \cite{DeY86}. The glassy properties
arise from the electric and elastic interaction between the CN$^-$
ions leading to an orientational disorder characterized by a broad
distribution of the parameters of the tunneling systems like in
structural glasses.

At low concentration ($x<0.001$) the interaction between the CN$^-$
ions can be neglected and the tunneling systems can be described in
terms of isolated defects. Because of the cubic symmetry of KBr,
eight potential minima in $\langle111\rangle$ direction exist for
isolated CN$^-$ ions. For the resulting tunneling splitting
${\it\Delta}_0/k_{\rm B}$ of the ground state values between 0.5 and
1.5\,K have been derived from measurements of specific heat
\cite{DeY86,Dob86}, thermal conductivity \cite{Sew66} and infrared
absorption \cite{Bey75,Var80}. At intermediate concentrations
($0.01<x<0.1$) the presence of pairs of strongly coupled CN$^-$ ions
at next nearest neighbor sites plays an important role
\cite{Lut82,Ens94,Ens95}. The tunnel splitting of such pairs is only
of the order of 10\,mK.

Until very recently it was the general belief that the dielectric
properties of insulating glasses -- free of magnetic impurities --
are largely independent of external magnetic fields. However, new
investigations have shown that the low-temperature dielectric
properties of certain multi-component glasses are extremely
sensitive to magnetic fields. In particular, the low-frequency
dielectric susceptibility \cite{Str98,Str00,Woh01,Hau01,Coc02,Hau02}
and the amplitude of spontaneous polarization echoes \cite{Lud02}
show a striking non-monotonic dependence on the applied magnetic
field.

To explain these findings it has been suggested that atomic
tunneling systems couple directly to magnetic fields
\cite{Ket99,Wue02}. In the model proposed by Kettemann {\it et
al.} it is assumed, that some tunneling particles exist that have
not just one path along which they can tunnel between the two
potential minima but several
--- like in a Mexican hat type of potential. The presence of a
magnetic field breaks the time reversal symmetry and thus changes
the tunneling probability along different paths. As a result the
tunnel splitting ${\it\Delta}_0$ becomes magnetic field dependent.
Since for a single tunneling system this effect is rather weak it
has been suggested that the coupling of a large number of such
tunneling systems leads to an enhancement of the magnetic field
dependence. The model by W\"urger considers pairs of weakly
interacting two-level systems consisting of systems with roughly the
same energy splitting. Such pairs have four levels, the middle two
of which are almost degenerate. Due to their interaction the
tunneling path of each individual tunneling system is slightly
deformed, effectively turning the tunneling path of such pairs into
a loop. Magnetic fields change the splitting of the almost
degenerate levels noticeably and influence in this way the
dielectric properties of the glass. W\"urger estimated that this
effect would be of the right order of magnitude to explain the
observed magnetic field dependence of the dielectric susceptibility
in multi-component glasses.

To further investigate the influence of magnetic fields on tunneling
systems, we have performed two-pulse polarization echo experiments.
Since in glasses the microscopic nature of the tunneling splitting
is hitherto unknown, we have decided to look for a possible magnetic
field effect in crystals with tunneling defects.
(KBr)$_{0.925}$(KCN)$_{0.075}$ was chosen because it is known to
show intense polarization echoes at low temperatures due to strongly
coupled pairs of CN$^-$ ions. The sample was placed in the uniform
field region of a re-entrant microwave cavity, loaded having a
resonance frequency of about $\omega_{\rm}/(2\pi)\approx 1$\,GHz.
Two short rf pulses, 100\,ns and 200\,ns in duration, separated by a
delay time $t_{12}$ of a few $\mu$s were used to generate
spontaneous polarization echoes occurring at $2t_{12}$. The
amplitude of such echoes is proportional to the number of resonant
tunneling systems that stay coherent during the time interval
$2t_{12}$, or in other words, have not undergone phase disturbing
processes.

\begin{figure}[t]
\includegraphics{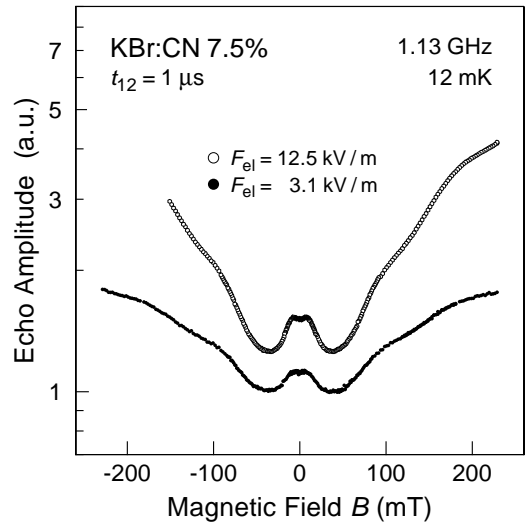}
\vskip -7mm \caption{Magnetic field dependence of the amplitude of
two-pulse echoes generated in
\hbox{(KBr)$_{0.925}$(KCN)$_{0.075}$} excited at two different
electric field strengths.} \label{fig1}
\end{figure}

Fig.~\ref{fig1} shows the magnetic field dependence of the amplitude
of two-pulse echoes generated in a single crystal
(KBr)$_{0.925}$(KCN)$_{0.075}$ at two different electrical field
strengths. As in glasses the echo amplitude is strongly influenced
by external magnetic fields leading to a non-monotonic variation.
Therefore, we conclude that the magnetic field dependence of the
dielectric response is not a unique property of certain glasses, but
also occurs in very different materials like single crystals
containing point defects but no magnetic impurities. Since the
origin of the tunneling states in crystals and their microscopic
parameters are well-known, studies of the magnetic field effects in
such materials should be very useful to gain insight into these
intriguing effects.

The data shown in Fig.~\ref{fig1} have been obtained at two
different amplitudes of the microwave pulses used to excite the
echo. Clearly, at the larger electric field amplitude the magnetic
effect is enhanced, indicating a non-linear electric field
dependence. The overall pattern, however, appears to be unaltered.
This observation agrees well with previous echo investigations of
a-BaO-Al$_2$O$_3$SiO$_2$. A closer inspection of the magnetic field
dependence reveals several distinct features. Close to zero field, a
tiny minimum appears, that was not observed in case of
a-BaO-Al$_2$O$_3$SiO$_2$. In addition, several bumps on the slope
towards higher magnetic fields are present. It is tempting to assign
the occurrence of these additional structures to the much narrower
distribution of the tunnel splitting of CN$^-$ pairs in
(KBr)$_{0.925}$(KCN)$_{0.075}$ compared to the distribution in
glasses. It might be that in amorphous solids the corresponding
features are simply washed out.

\begin{figure}[t]
\includegraphics{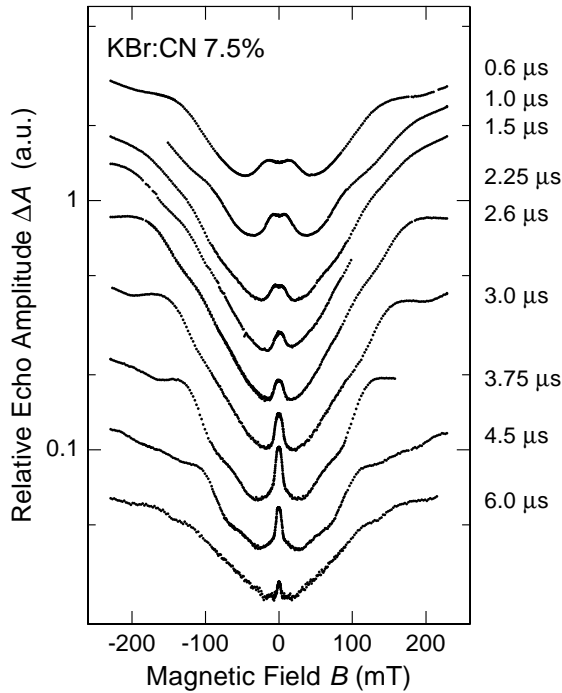}
\vskip -7mm \caption{Amplitude of two-pulse echoes generated in
(KBr)$_{0.925}$(KCN)$_{0.075}$ with different delay times as a
function of the applied magnetic field. The delay times are
indicated on the left-hand side of this figure.} \label{fig2}
\end{figure}

\begin{figure}[t]
\includegraphics{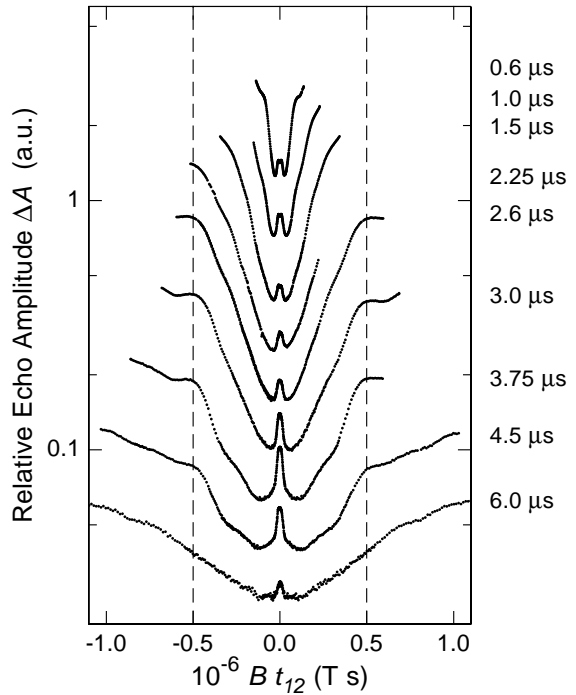}
\vskip -7mm \caption{Amplitude of two-pulse echoes generated in
(KBr)$_{0.925}$(KCN)$_{0.075}$ as a function of the product of the
applied magnetic field and the delay time for different delay
times $t_{12}$. The two dashed lines mark the position of
features, which appear at identical values of $Bt_{12}$.}
\label{fig3}
\end{figure}

Surprisingly, the pattern changes with a variation of the delay time
between the pulses. Fig.~\ref{fig2} shows the magnetic field
dependence of the amplitude of two-pulse echoes in
(KBr)$_{0.925}$(KCN)$_{0.075}$ obtained at different delay times.
Obviously there is a systematic relation between the delay time and
the magnetic field at which certain features occur. Note in
particular the broadening of the central peak with decreasing delay
time. This dependence indicates that the magnetic field has an
influence on the phase evolution of the resonant tunneling systems,
because between the two excitation pulses as well as between the
second pulse and the echo the phase develops freely. To strengthen
this point further we have plotted in Fig.~\ref{fig3} the data of
Fig.~\ref{fig2} as a function of the product $Bt_{12}$ of magnetic
field~$B$ and delay time~$t_{12}$. Clearly, the feature at $B\approx
0$ has now the same width for all curves and the additional
structure seen at higher fields appears at roughly the same
$Bt_{12}=0.5\times10^{-6}\,$Ts as indicated by the two dashed lines
in Fig.~\ref{fig3}.

We conclude from this observation that the magnetic field couples
to the tunneling states and changes the phase linearly with time.
An obvious problem of this interpretation is that the second pulse
applied in our experiment causes formally a time reversal of the
phase development and therefore any additional phase shift
accumulated during $t_{12}$ should be compensated for in the time
interval from $t_{12}$ to $2t_{12}$. Since this is obviously not
the case we have to assume that the magnetic field breaks the time
reversal symmetry in this experiment.

One possible scenario that could lead to such a symmetry breaking
would be a change of the magnetic moment of the tunneling systems by
the microwave pulses. This is conceivable if the tunneling systems
are not spherically symmetric and therefore neither the ground state
nor the excited states are eigenstates of the angular momentum
operator. Although isolated CN$^-$ tunneling states in KBr crystals
should exhibit a spherical symmetry, at a concentration of 7.5\% the
elastic interaction between the defects lead to an orientational
disorder and therefore in general to a non-spherical symmetry of the
tunneling systems.

It should be added that our experiments show that we are dealing
with asymmetric tunneling states, because the tunnel splitting of
the relevant pairs is about 10\,mK \cite{Ens94} and the experiments
have been performed at a resonance energy of 1.13\,GHz corresponding
to an energy of 56\,mK. For such asymmetric tunneling states the
effective magnetic moment will be different for each eigenstate.
This means that the phase change accumulated during the
time~$t_{12}$ before the inverting pulse and afterwards do not
cancel exactly when the echo appears. This effect could lead to an
enhanced or reduced polarization at $2t_{12}$ depending on the delay
time in the experiment. Perhaps this scenario is not the only
possible way in which the magnetic field breaks the time inversion
symmetry in this experiment. At the present time, however, no other
mechanism is known.

Let us now analyze more quantitatively the features -- minima and
maxima -- , that coincide on the $Bt_{12}$ plot. The width of the
central peak of the curves in Fig.~\ref{fig3} is about $0.2\times
10^{-6}\,$Ts, meaning the corresponding minimum appears at $\vert
Bt_{12}\vert=0.1\times 10^{-6}\,$Ts. The unit of this quantity is
mass divided by charge. If we assume that charges comparable with an
elementary charge are involved we may conclude that a mass of the
order of 10 proton masses is required to account for this minimum,
or equivalently, a magnetic moment of the order of a nuclear
magnetic moment. The features indicated by the two dashed lines
appear roughly at $\vert Bt_{12}\vert=0.5\times 10^{-6}\,$Ts,
indicating that a 5 times larger mass or in turn a 5 times smaller
magnetic moment seems to be involved. At the present time this
analysis is completely empirical, because we do not know whether the
magnetic moments observed in our experiment are due to the orbital
motion of tunneling particles or due to the nuclear magnetic moments
associated with the CN$^-$ molecule. In the latter case it would by
unclear why the nuclear magnetic moment is so strongly coupled, to
the tunneling motion of the CN$^-$ ions.

The overall increase of the echo amplitude towards larger fields
does not depend on the delay time and therefore cannot be rescaled
in a $Bt_{12}$-plot. It seems that this effect is caused by a
different mechanism and is not directly related to the occurrence of
the specific features mentioned above. If we again cast a glance at
Fig.~\ref{fig1} we see that the curve obtained at higher excitation
amplitude shows an increase of the echo amplitude at the largest
magnetic field of 230\,mT of about a factor of 3 compared to the
zero field data, with the tendency of a further increase. It seems
that roughly a factor of 3 more tunneling systems are moving in
phase coherently at 230\,mT. Since the zero field data have been
obtained after adjusting the experimental parameters such that the
echo amplitude had its maximum for the applied excitation pulses,
the strong increase of the echo amplitude with magnetic field
appears to be very mysterious.

Two possibilities for an explanation are conceivable: Either
magnetic fields suppress an absorption mechanism that reduces the
observed echo amplitude at zero field, or magnetic fields suppress
phase perturbing processes, which dominate the dephasing a zero
magnetic field. Studies of the microwave absorption in magnetic
fields have shown no indication for a magnetic field dependent
absorption mechanism, ruling out the first of the two possibilities.
In order to investigate the second explanation, we have performed
measurements of the echo decay as a function of $t_{12}$ in
different applied magnetic fields. Only a very slight influence of
the magnetic field on the decay of the echo is found, which is by
far not sufficient to explain the factor of three increase of the
echo amplitude in magnetic fields. However, the measurement of the
echo decay as a function of the delay time is only sensitive to
processes which influence the free development of the phase between
and after the two excitation pulses. Any phase disturbing process
that would act only during the excitation pulses would not change
the decay pattern as a function of delay time and therefore cannot
be ruled out as a possibility. We have no plausible physical picture
how such a dephasing process limited to the excitation pulses can
come about, but the strong dependence of the magnetic field effect
on the amplitude of the microwave field might be taken as a support
for the validity of this interpretation.

In summary, magnetic field effects of the dielectric properties have
been observed for the first time for a crystalline system containing
point defects. A strong non-monotonic increase of the amplitude of
spontaneous polarization echoes with applied magnetic field has been
fund in (KBr)$_{0.925}$(KCN)$_{0.075}$. Certain features in the
variation of the echo amplitude with the magnetic field depend on
the delay time between the two exciting pulses. This observation
clearly indicates that the external magnetic field couples to the
tunneling systems and changes their phase relative to the external
electric field in a well-defined way. The overall increase of the
echo amplitude and the dependence on the electrical field strength
might be explained by a hitherto unknown dephasing mechanism that
only occurs during the excitation pulses.

We thank R. Weis, A. W\"urger, M.v. Schickfus, P. Strehlow and S.
Hunklinger for helpful comments and experimental support. We are
grateful to F. Luty for providing the sample. This work was
supported by the Deutsche Forschungsgemeinschaft (Grant No.
Hu359/11).


\end{document}